\documentclass[structabstract]{aa}  
\usepackage{graphicx}
\usepackage{txfonts}
\usepackage{natbib}
\bibpunct{(}{)}{;}{a}{}{,} % to follow the A&A style
\usepackage{color}
\newcommand{\cmtwo}{cm$^{-2}$}  
\newcommand{\cmthree}{cm$^{-3}$}

\newcommand{\kms}{km\,s$^{-1}$}       %km/s
 
\newcommand{\vlsr}{$\upsilon_{\rm LSR}$}        %velocities
\newcommand{\dv}{$\Delta \upsilon_{\rm FWHM}$}

\newcommand{\tas}{$T^{*}_{\rm A}$}
\newcommand{\tadv}{$\int \!\! T^{*}_{\rm A} {\rm d}\upsilon$}
\newcommand{\tmb}{$T_{\rm mb}$}
\newcommand{\trms}{$T_{\rm rms}$}
\newcommand{\tsys}{$T_{\rm sys}$}
   
\newcommand{\trs}{$T^{*}_{\rm R}$}

\newcommand{\um}{$\mu$m}                                 %micron
\newcommand{\molh}{H$_{2}$}                              %H_2, H_2O, H II

\newcommand{\water}{H$_{2}$O}

\newcommand{\molo}{O$_{2}$}			% O_2, 16O18O, C18O, 13C18O
\newcommand{\moloiso}{${\rm O}^{18}{\rm O}$}
\newcommand{\moloisoiso}{$^{16}{\rm O}^{18}{\rm O}$}
\newcommand{\moloisotwo}{$^{16}{\rm O}^{18}{\rm O}\,(2_1-0_1)$}
\newcommand{\coiso}{${\rm C}^{18}{\rm O}$}
\newcommand{\coisoiso}{$^{13}{\rm C}^{18}{\rm O}$}
\newcommand{\msun}{$M_{\odot}$}

\newcommand{\gapprox}{$\stackrel {>}{_{\sim}}$}   %greater/less approx.
\newcommand{\lapprox}{$\stackrel {<}{_{\sim}}$}
\newcommand{\about}{$\sim$}                       %approx
\newcommand{\powten}[1]{10$^{#1}$}
\newcommand{\ntwohp}{${\rm N_2H^+}$}
\newcommand{\av}{$A_{\rm V}$}                     %extinction

\newcommand{\ro}{$\rho \, {\rm Oph}$}
\newcommand{\roc}{$\rho \, {\rm Oph \,\, cloud}$}
\newcommand{\roa}{$\rho \, {\rm Oph \, A}$}

\newcommand{\roac}{$\rho \, {\rm Oph \, A \, core}$}
\newcommand{\roca}{$\rho \, {\rm Oph \, A \, cloud}$}

\newcommand{\amin}{$^{\prime}$}                   %arcus and coordinates
\newcommand{\asec}{$^{\prime \prime}$}

\newcommand{\rahms}[3]{\mbox{#1$^{\rm h}$#2$^{\rm m}$#3$^{\rm s}$}}
\newcommand{\radot}[4]{\mbox{#1$^{\rm h}$#2$^{\rm m}$#3$\stackrel {\rm s}{_{\bf\cdot}}$#4}}  
\newcommand{\decdms}[3]{\mbox{#1$^{\circ}$#2$^{\prime}$#3$^{\prime \prime}$}}

\newcommand{\asecdot}[2]{\mbox{#1$\stackrel {\prime \prime}{_{\bf \cdot}}$#2}}

\begin{document}
   \title{\moloiso\ and \coiso\ observations of \roa\thanks{Based on observations with APEX, Llano Chajnantor, Chile.}}

%   \subtitle{}

    \author{R. Liseau\inst{1}  
       	 \and
          B. Larsson\inst{2}
          \and
          P. Bergman\inst{1}
           \and 
          L. Pagani\inst{3}
           \and
          J.H. Black\inst{1}
          \and
          \AA. Hjalmarson\inst{1}
           \and
           K. Justtanont\inst{1}
	 }

   \offprints{R. Liseau}

   \institute{Department of Radio and Space Science, Chalmers University of Technology, Onsala Space Observatory, SE-439 92 Onsala, Sweden,
              \email{firstname.familyname@chalmers.se}
 	\and
 	Department of Astronomy, Stockholm University, AlbaNova, SE-106 91 Stockholm, Sweden, \email{bem@astro.su.se}
 	\and
 	LERMA, L'Observatoire de Paris, 61, avenue de l'Observatoire, F-75014 Paris, France,  \email{laurent.pagani@obspm.fr}	
	}

   \date{Received ; accepted }

% \abstract{}{}{}{}{} 
% 5 {} token are mandatory

  \abstract
   {Contrary to theoretical expectation, surprisingly low concentrations of molecular oxygen, \molo, have been found  in the interstellar medium by means of orbiting telescopes.}
   {Observations of the $(N_J=1_1-1_0)$ ground state transition of \molo\ with the Odin satellite resulted in a \gapprox\,$5 \sigma$ detection toward the dense core \roa. At the frequency of the line, 119\,GHz, the Odin telescope has a beam width of 10\amin, larger than the size of the dense core, so that the precise nature of the emitting source and its exact location and extent are unknown. The current investigation is  intended to remedy this.}
   {Telluric absorption makes ground based \molo\ observations essentially impossible and observations had to be done from space. Millimetre-wave telescopes on space platforms were necessarily small, which resulted in large, several arcminutes wide, beam patterns. Although the Earth's atmosphere is entirely opaque to low-lying \molo\ transitions, it  allows ground based observations of the much rarer \moloisoiso\ in favourable conditions and at much higher angular resolution with larger telescopes. In addition, \roa\ exhibits both multiple radial velocity systems and considerable velocity gradients. Extensive mapping of the region in the proxy \coiso\ $(J=3-2)$ line can be expected to help identify the \molo\ source on the basis of its line shape and Doppler velocity. Line opacities were determined from observations of optically thin \coisoiso\ $(J=3-2)$ at selected positions.}
   {During several observing periods, two \coiso\ intensity maxima in \roa\ were searched for in the \moloisotwo\ line at 234\,GHz with the 12\,m APEX telescope. These positions are associated with peaks in the mm-continuum emission from dust.  Our observations resulted in an upper limit on the integrated \moloiso\ intensity of \tadv\,$ < 0.01$\,K\,\kms\ ($3 \sigma$) into the \asecdot{26}{5} beam.} 
   {Examining the evidence, which is based primarily on observations in lines of \moloiso\ and \coiso, leads us to conclude that the source of observed \molo\ emission is most likely confined to the central regions of the \roac. In this limited area, implied \molo\ abundances could thus be higher than previously reported, by up to two orders of magnitude.}  

   \maketitle
%
%________________________________________________________________

\section{Introduction}

Oxygen is the most abundant of the astronomical metals \citep[e.g.,][and~references~therein]{asplund2009}. Consequently, in its molecular form, it was also expected to be very abundant in the UV-shielded regions inside molecular clouds \citep[e.g.,][]{black1984,bergin2000,charnley2001,roberts2002,spaans2001,viti2001,willacy2002,quan2008} and to contribute significantly to the cooling, hence the energy balance, of dense clouds \citep{goldsmith1978}.

Because of the high \molo\ content in the Earth's atmosphere, astronomical \molo\ sources cannot be observed from the ground. Dedicated space missions\footnote{ SWAS in 1998, see {\ttfamily http://cfa-www.harvard.edu/swas/}, and Odin in 2001, see {\ttfamily http://www.snsb.se/eng\_odin\_intro.shtml} } came into operation near the beginning of the new millenium. Their unsuccessful searches \citep{goldsmith2000,pagani2003} were highly disappointing and it was hard to understand that, in the interstellar medium (ISM), \molo\ is an elusive species (see references cited above).

Eventually, after more than 20 days of Odin-observing during three different runs, came a real break-through: for the very first time, \molo\ was finally detected in the ISM \citep{larsson2007}. The \molo\ emitting object, \roa, is a dense clump \citep{loren1990} in a region of active star formation (L\,1688). On the basis of theoretical model calculations, the
detection on the basis of theoretical model calculationsof this kind of source had earlier been predicted by \citet{black1984} and \citet{marechal1997a}, where the latter authors made their specific prediction with regard to Odin.

Odin carries a 1.1\,m telescope which is designed for observations in the submillimetre regime, between roughly 480 and 580\,GHz ($0.5-0.6$\,mm). However, the \molo\ discovery was made with a dedicated 119\,GHz (2.5\,mm) receiver aboard Odin, fix-tuned to the frequency of the ground state O$_2$ ($N_J = 1_1 - 1_0$) transition at 118\,750.343\,MHz. At this frequency, the telescope beam size is 10\amin, larger than the angular dimension of the dense \roac, which is about 4\amin\ \citep[FWHM of devonvolved CS core,][]{liseau1995}. 

It follows that the true \molo\ source is likely under-resolved, the consequence of which directly affects estimates of the abundance of \molo, i.e. $N$(\molo)/$N$(\molh): depending on the adopted model, the Odin observations imply an abundance which is currently uncertain by two orders of magnitude \citep{liseau2005}. 

In Fig.\,2 of \citet{larsson2007}, the Odin-\molo\ line is compared to transitions of other molecular species in \roa. Whereas lines of \water\ and CO are optically very thick over large parts of the cloud and have self-absorbed profiles, the optically thin \molo\ line displays a simple, Gaussian shape. This line shape is similar to that of a carbon recombination line, displayed at the top of the figure and which most likely originates in the \ro-PDR. If also the main source of \molo\ emission, the abundance would indeed be very low.

However, the \molo\ line shape is also similar to that of the C$^{18}$O (3-2) line, also shown in the figure. This suggests that the C$^{18}$O line can be used as a tracer of the molecular oxygen emission and we set out to map the 10\amin\ Odin beam in the (3-2) transition of C$^{18}$O with the APEX beam of size 19\asec.  It was expected that a detailed comparison of the line centre velocity with that of the \molo\ line would help to narrow down the exact location of the \molo\ emission, since two distinct velocity components are known to be present in \roa. This information is needed to understand, where, i.e. in what physical conditions, the majority of the \molo\ molecules is excited: in the cold and dense dark cores \citep{difrancesco2004}, in the extended warm Photon Dominated Region \citep[PDR;][]{hollenbach2009} or in the hot shocked gas of the outflow from VLA\,1623 \citep{liseau2009}? With the \coiso\ proxy for \molo\ emission, probable emission regions were identified, which were then observed for \moloisotwo. 

There exists earlier work for this line and the \roc. \citet{goldsmith1985} observed \roa\ in the same transition and with comparable beam size (26\asec), albeit at an offset 11\asec\,E and 61\asec\,N relative to the position of SM\,1N. They obtained  \trs$ <120$\,mK ($1 \sigma$) over 0.34\,\kms. At similar channel resolution (0.32\,\kms) and toward essentially the same position, \citet{liszt1985} obtained an rms-noise value of \trs\,$<17.5$\,mK with the 12\,m NRAO telescope (34\asec). These papers also present energy level diagrams. Observations made with the 10\,m telescope of the Caltech Submillimeter Observatory (CSO) in July 1991 and for the position \rahms{16}{23}{25}, \decdms{$-24$}{15}{49} (B1950) \footnote{This corresponds to \radot{16}{26}{26}{4}, \decdms{$-24$}{22}{33} in J2000 coordinates and is at (+25\asec, +80\asec) relative to the origin of the \coiso\ map (Fig.\,\ref{c18o32_map}).}  resulted in an rms noise temperature of 16\,mK in a 0.25\,\kms\ velocity bin and of 12\,mK after binning to 0.50\,\kms\ (E. van Dishoeck, J. Keene \& T. Phillips, private communication). 

The derivation of molecular abundances requires knowledge of the \molh\ column density. One of the widely exploited techniques to estimate $N$(\molh) is to use observations of C$^{18}$O, the transitions of which in many cases can be shown to be optically thin. We discovered, however, that in the dense core regions of \roa, this not to be the case everywhere and that appropriate opacity corrections using the $^{13}$C$^{18}$O line needed to be made. 

This paper is organised as follows: In Sect.\,2, our APEX observations of the \roca\ in transitions of \moloiso, \coiso\ and \coisoiso\ are described. Sect.\,3 presents our results, which are discussed in Sect.\,4. Finally, in Sect.\,5 our main conclusions are briefly summarised.

\section{Observations and data reductions}

All observations have been made with the SIS receivers and spectrometers at the Atacama Pathfinder EXperiment (APEX). The 12\,m APEX telescope is located at an altitude of about 5100\,m on the Llano de Chajnantor in northern Chile\footnote{\ttfamily http://www.apex-telescope.org/}. The telescope pointing is accurate to 3\asec\ (rms).

The Fast Fourier Transform Spectrometer (FFTS) was configured to have 8192 channels, which over a bandwidth of 1\,GHz provides a resolution of 122\,kHz, corresponding to 0.16\,\kms\ and 0.11\,\kms\ at 234\,GHz and 329\,GHz, respectively. As frontends for these frequencies, we used APEX\,1 of the Swedish Heterodyne Facility Instrument \citep[SHFI,][]{vassilev2008} and APEX\,2A \citep{risacher2006}.

\subsection{\moloiso\ observations}

The data have been collected during three different observing runs in 2008 and 2009. The frequency of the ($2_1-0_1$) line can be derived from the data given by \citet{steinbach1975} as 233946.179\,MHz. At 234\,GHz, the APEX beam has a half power beam width HPBW=\asecdot{26}{5} and the main beam efficiency is $\eta_{\rm mb}=0.75$. The telescope was pointed toward RA=\radot{16}{26}{27}{2} and Dec=\decdms{$-24$}{23}{34} (J2000), a position which was initially chosen on the basis of, as it turned out, insufficiently sampled data (see Sect.\,3.2). In addition, the strongest peak of doubly deuterated formaldhyde emission in the \roac\ (P.\,Bergman et al., in preparation)\footnote{The 234\,GHz spectra admitted also lines of deuterated formaldehyde. Mapping observations revealed this peak position.}, which is situated 30\asec\ south of these coordinates, was also observed. These positions are close to the location of intense mm-dust-emission (cf. Fig.\,\ref{c18o32_map}), i.e. the dense core SM\,1 \citep{motte1998}. For the primary position, the total on-source integration time was 4.9 hours and the average system temperature was \tsys\,\about\,220\,K, whereas for the $-30$\asec-position, these values were 6.5 hours and 210\,K, respectively.

\subsection{\coiso\ and \coisoiso\ observations}

The observations were collected during two observing runs in 2006 and 2007 at the APEX telescope. The observing mode was position switched raster mapping and the data were sampled according to the Nyquist criterion on a rectangular 10\asec\ grid, aligned with the equatorial coordinate system ($200^{\prime \prime} \times 200^{\prime \prime}$). At 329\,GHz, the HPBW=19\asec\ and the average system temperature was \tsys=200\,K. The efficiencies were $\eta_{\rm mb}=0.73$ and $\eta_{\rm Moon}=0.85$ for point source and extended source calibrations, respectively. 

In addition, an extended raster map of the outer regions of \roa\ was obtained on a coarser grid with 20\asec\ (full beam) spacings. The entire region observed is thus as large as $\Delta \alpha \times \Delta \delta = 10^{\prime} \times 5^{\prime}$.   

The origin of the map is the same as that of the Odin observations, i.e. the (0, 0) position is at RA=\radot{16}{26}{24}{6} and Dec=\decdms{$-24$}{23}{54} (J2000). The same reference position as for the Odin observations \citep{larsson2007}, viz. 15\amin\,N relative to the map centre, was used here for calibration purposes. In addition to the \coiso\ map, five positions were also observed in the (3-2) transition of the even rarer isotope \coisoiso\ (Table\,\ref{13c18o32}).  \citet{klapper2003} provide lab-frequencies for the (3-2) rotational transition of \coiso\ and \coisoiso, i.e.,  329\,330.552\,MHz and 314\,119.660\,MHz, respectively, and where the latter is a weighted mean value, with the $^{13}$C hyperfine structure being ignored. 
 
\begin{figure}
  \resizebox{\hsize}{!}{
  \rotatebox{00}{\includegraphics{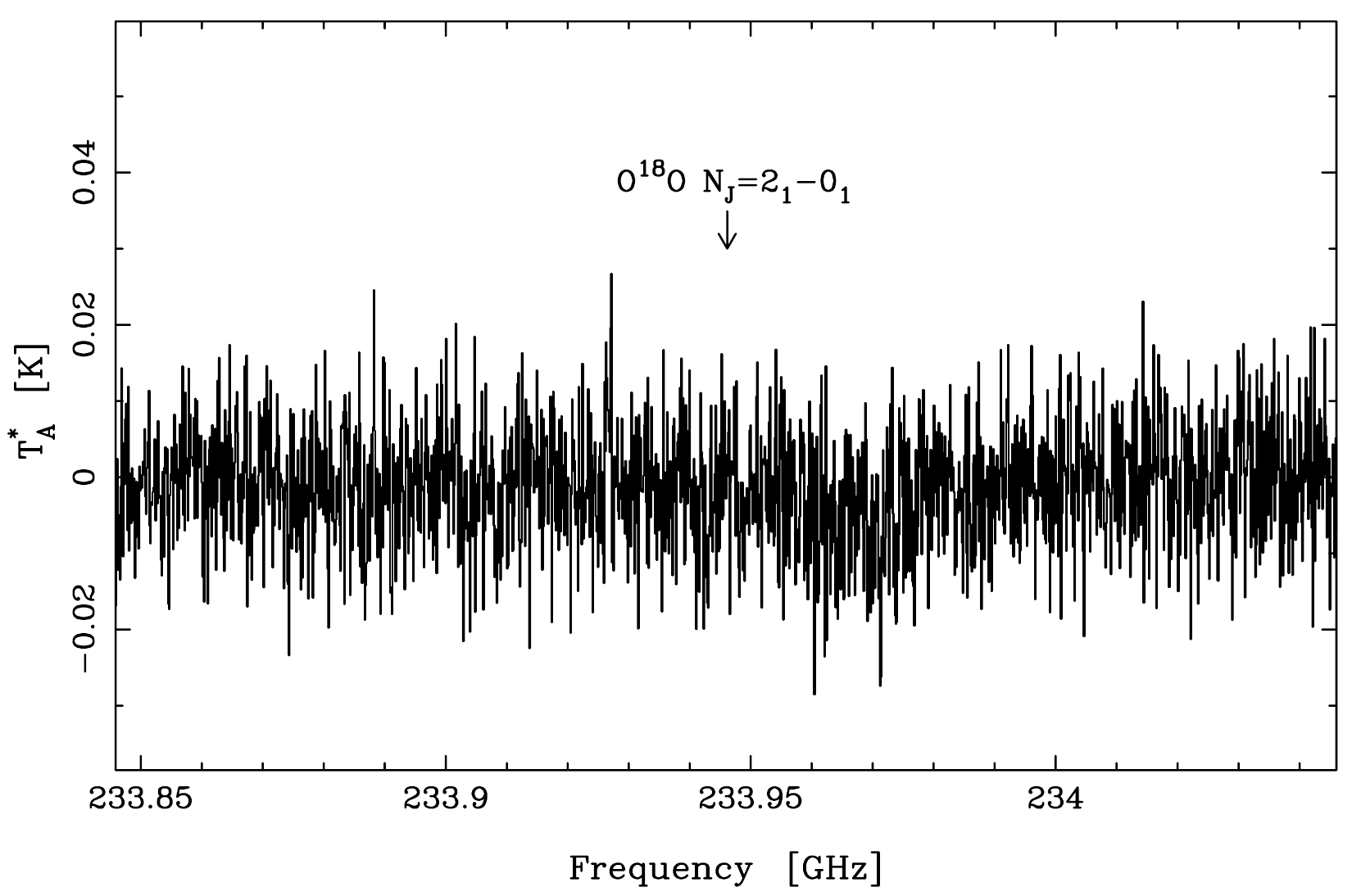}}
  }
  \caption{The central part of the 1\,GHz wide APEX spectrum centered on the frequency of the \moloisotwo\ transition, 233.946179\,GHz, and obtained toward RA=\radot{16}{26}{27}{2} and Dec=\decdms{$-24$}{24}{04} (J2000) in \roa. The sampling is in 122\,kHz wide channels ($\delta \upsilon = 0.16$\,\kms).
   }
  \label{spectrum}
\end{figure}

\begin{figure}
 \resizebox{\hsize}{!}{
   \rotatebox{00}{\includegraphics{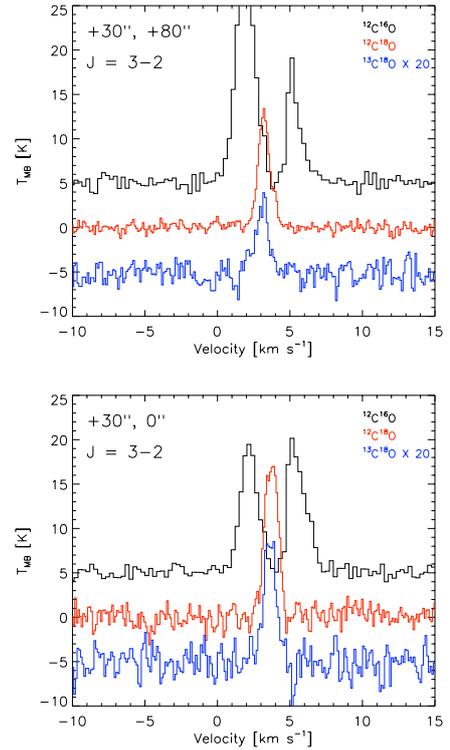}}
   }
   \caption{($J=3-2$) spectra (\tmb\ vs \vlsr) of, from top to bottom, CO (black), \coiso\ (red) and \coisoiso\ (blue) toward two positions in the \roac \ (cf. Fig.\,\ref{c18o32_map}). For clarity, two of the spectra are offset by $\pm 5$\,K and the \coisoiso\ spectra have been multiplied by a factor of twenty.
    }
   \label{co_lines}
 \end{figure}

\section{Results}

\subsection{\moloiso}

\begin{figure}[ht]
  \resizebox{\hsize}{!}{
     \rotatebox{00}{\includegraphics{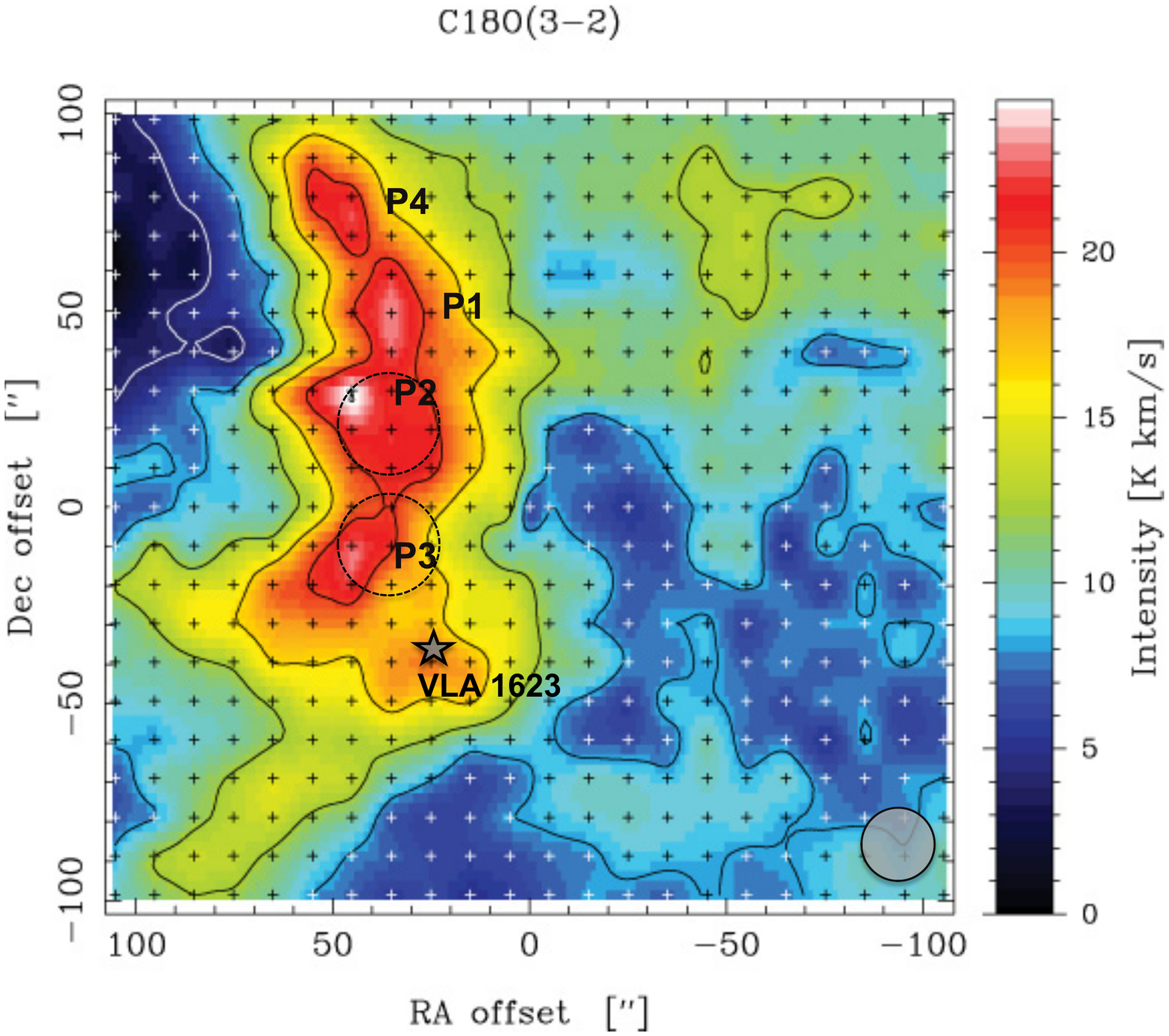}}
  }
  \caption{\coiso\,(3-2) integrated intensity, \tadv, of the dark core \roa. The map was obtained with APEX and observed positions are shown as crosses. The beam size at 329\,GHz is shown in the lower right corner. Offsets are with respect to the origin, RA=\radot{16}{26}{24}{6} and Dec=\decdms{$-24$}{23}{54} (J2000). The position of the outflow driving Class\,0 source VLA\,1623 is shown by the star symbol. P\,1-P\,4 designate the clumps discussed in the text (Table\,\ref{positions}). The beam size at 234\,GHz is indicated by the dotted circles, at the observed \moloiso\ positions.
   }
  \label{c18o32_map}
\end{figure}

The  \moloisotwo\ line was not detected toward any of the observed positions. Toward the position associated with P\,2 (see Fig.\,\ref{c18o32_map} and Table\,\ref{positions}), the noise level is \trms=6.5\,mK ($1 \sigma$) in a 0.62\,\kms\ bin. The result is similar for the observation of the position 30\asec\ south (P\,3), i.e., \trms=8.2\,mK in a 0.16\,\kms\ bin (Fig.\,\ref{spectrum}).

\begin{table}[t]
\begin{flushleft}
 \caption{\label{positions} \coiso-peaks of integrated intensity, \tadv}
\resizebox{\hsize}{!}{
\begin{tabular}{lllll}
\hline 
\noalign{\smallskip}
\coiso-peak &	Offset (arcsec)	& R.A. (J2000) & Dec. (J2000) & Other ID  \\             
\noalign{\smallskip}
\hline
  \hline
  \noalign{\smallskip}
P\,1	& +35, +50	& 16:26:27.2	& $-24$:23:04	& N\,1		\\
P\,2	& +45, +28	& 16:26:27.9	& $-24$:23:26	& N\,5, SM\,1N	\\
P\,3	& +45, $-13$	& 16:26:27.9	& $-24$:24:07	& N\,4, SM\,1	\\
P\,4	& +48, +75	& 16:26:28.1	& $-24$:22:39	& 16264-2422b\\
 \noalign{\smallskip}
  \hline
  \end{tabular}
    }
\end{flushleft}
  Note to the Table: N-sources from \citet{difrancesco2004}, SM-objects from \citet{motte1998}, RA-Dec labelled source from \citet{johnstone2000}.
\end{table}

\subsection{\coiso\ and \coisoiso}

Example spectra in three isotopes of CO (3-2) are shown in Fig.\,\ref{co_lines} toward two positions in the central region of the \roac. Further, Fig.\,\ref{c18o32_map} shows the inner, high-resolution, map of integrated intensity, \tadv, of the \coiso\,(3-2) line. Within a range of R.A. offsets +30\asec\ to +50\asec, four distinct intensity peaks are discernable. In Table\,\ref{positions}, these are designated P\,1 through P\,4 and their J2000 coordinates are given. The C$^{18}$O line is very narrow, e.g. merely 1.0\,\kms\ (FWHM) at the inconspicuous (0, 0) position. 

\begin{figure*}[ht]
  \resizebox{\hsize}{!}{
  \rotatebox{00}{\includegraphics{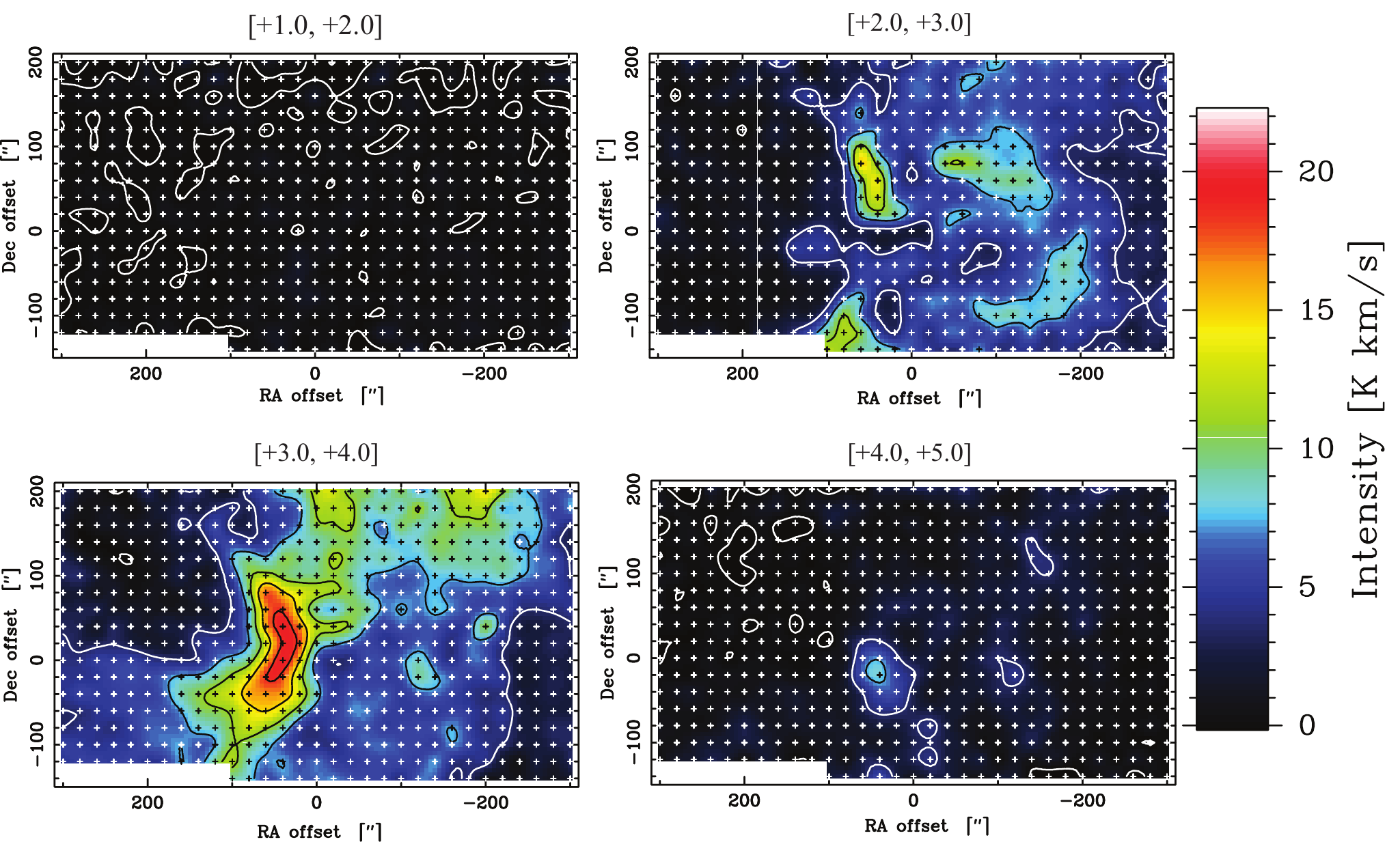}}
  }
  \caption{A mosaic of maps of integrated \coiso\,(3-2) line intensity over 1.0\,\kms\ wide velocity intervals, from +1.0 to +5.0\,\kms. The extended maps with 20\asec\ spacing are shown. Two velocity components are identified in the \roac, falling into the [2, 3] and [3, 4] bins, respectively. The lowest contour level corresponds to 4\,K\,\kms\ and increments are also by this amount.
   }
  \label{c18o32_mosaic}
\end{figure*}

Examination of the entire data set for \coiso\,(3-2) reveals the fact that, within the mapped region, maximum emission occurs at LSR-velocities +2.7 to +3.7\,\kms. This velocity interval corresponds to that of the \molo\,119\,GHz emission, viz. \vlsr\,\about\,+2.5 to +3.5\,\kms\ \citep{larsson2007} and Fig.\,\ref{convolved} here. \roa\ displays a complex velocity field and two distinct velocity components can be identified, giving rise to spectral line blending. These components are essentially confined within the LSR-velocity bins [+2, +3] and [+3, +4] (in \kms). Fig.\,\ref{c18o32_mosaic} presents a mosaic of the integrated line intensity in 1.0\,\kms\ wide bins. A variety of this kind of channel maps demonstrates quite convincingly that the location of the \molo\ emitting gas is most likely associated with the central core region of \roa. 

\begin{table}
\begin{flushleft}
 \caption{\label{13c18o32} Observed positions of \coisoiso\,(3-2) and line opacities, $\tau_{^{13}{\rm C}^{18}{\rm O}}$.}
\resizebox{\hsize}{!}{
\begin{tabular}{cccclll}
\hline 
\noalign{\smallskip}
Offset 			& \vlsr  	& FWHM 	& $T_{\rm peak}(^{12}{\rm C}^{18}{\rm O}),$ 	&  $\tau_{^{13}{\rm C}^{18}{\rm O}}$ & Note \\   
(arcsec)                       &  (\kms)   &    (\kms) & $T_{\rm peak}(^{13}{\rm C}^{18}{\rm O})$ (K)	&                                  &           \\ 
\noalign{\smallskip}
\hline
  \hline
  \noalign{\smallskip}
 $-60$, +80		& $2.86 \pm 0.05$	& $0.78 \pm 0.12$	& 10.5, 0.25			& 0.02	& $\ldots$			\\
   0, +60			& $3.13 \pm 0.06$	& $0.92 \pm 0.14$	&  \phantom{1}8.1, 0.22	& 0.03	& $\ldots$			\\
  +30, +80			& $3.09 \pm 0.05$	& $1.11 \pm 0.09$	& 13.0, 0.38			& 0.03	& P\,1			\\
  +30, 0			& $3.62 \pm 0.02$	& $1.02 \pm 0.05$	& 16.5, 0.71			& 0.04	& P\,3 			\\
  +60, $-100$		& $2.95 \pm 0.06$	& $1.17 \pm 0.15$	&  \phantom{1}6.1, 0.23	& 0.04	& 2 lines?			\\
 \noalign{\smallskip}
  \hline
  \end{tabular}
    }
\end{flushleft}
  Note to the Table: Error on line ratios is estimated at 15- 20\%.
\end{table}

\section{Discussion}

\subsection{The dense clumps of \roa}

The \coiso\ intensity maxima in Fig.\,\ref{c18o32_map} seem comparable in size with the APEX beam, which could indicate that the diameter of these clumps does not exceed 20\asec. From the comparison of their locations with those observed in the emission of the dust at 1.3\,mm \citep{motte1998} and 850\,\um\ \citep{johnstone2000} and of the quiescent gas in the \ntwohp\,(1-0) line \citep{difrancesco2004}, it becomes evident that P\,4 lacks correspondence with features at 1.3\,mm and \ntwohp\ emission, but shows up weakly at 850\,\um.  P\,1 likely is N1 (which is not seen in the dust maps), P\,2 corresponds to N\,5 (also prominent in the dust as SM\,1N), and P\,3 seems associated with N\,4 and SM\,1 \citep[also 16264-2423 of][]{johnstone2000}. Derived temperatures and densities for these clumps are of the order of 15-30\,K and $0.2 - 5 \times 10^6$\,\cmthree, respectively \citep[e.g.,][]{andre1993,motte1998,johnstone2000}.

In summary,  the evidence points toward the fact that also \molo\ is concentrated in the dense dark core regions, where the molecules would be protected against photo-dissociation due to the intense UV field (G$_0$ of the order of \powten{2}) generated by the two B-stars, east and west of the cores, respectively \citep{liseau1999}. The size of the \molo\ emitting regions appears not to exceed one arcminute, so that a conservative estimate of the Odin beam filling would be about  0.01. If the emission originates in a core of size \about\,20\asec\ or smaller, the Odin beam filling factor would be reduced by yet another order of magnitude. The \molo\ abundance would scale accordingly and could in this case be locally as high as a few times \powten{-5}, which would be comparable to the total abundance of oxygen in the gaseous phase \citep[e.g.,][]{liseau2009}. 

\subsubsection{Line optical depths}

The ratio of the \coisoiso\ and $^{12}$\coiso\ line intensities can be used to estimate the optical depth in the rarer isotope line,  $\tau_{^{13}{\rm C^{18}\rm O}}  \sim \ln {(1-r_{13})^{-1}}  $, where $r_{13} \equiv T(^{13}{\rm C}^{18}{\rm O})/T(^{12}{\rm C}^{18}{\rm O})$. From the data presented in Table\,\ref{13c18o32}, it is clear that the \coiso\,(3-2) line could have significant opacity along several lines of sight, unless the relative abundance [$^{12}$\coiso /\coisoiso] $\ll 50$ (or the excitation temperatures for these species differ substantially). 

\citet{federman2003} determined a column density ratio $N(^{12}{\rm CO})/N(^{13}{\rm CO}) = 125\pm 23$ toward a line of sight designated \roa\ by them\footnote{In addition, for their \roa\ line of sight, \citet{federman2003} also give $N(^{12}{\rm C^{16}O})/N(^{12}{\rm C^{18}O}) = 1100\pm 600$.}. However, their coordinates refer to the star, one degree north-north-west from the \roac\  discussed in our paper. In the associated nebula, the physical conditions are different from those in the dense core, possibly leading to different isotopic abundances. In the shielded regions of the dense cores, chemical isotopic fractionation can be expected to be of minor importance. It is worth noting that in the nearby  \ro\,core C, a lower isotopic ratio $[^{12}{\rm C}]/[^{13}{\rm C}]=65 \pm 10$ has been derived by \cite{bensch2001}.

\subsection{Column densities}

If local thermal equilibrium (LTE) is a good approximation for the level populations, the column density of all molecules of the species, $N$(mol) in \cmtwo, can be estimated from the observed intensity of an optically thin line,  viz.

\begin{equation}
N({\rm mol})  =   \frac { \int \!\! T^{*}_{\rm A} {\rm d}\upsilon } { f_{\rm b}\,\eta_{\rm mb} } \times \Phi(T_{\rm k}) 
\end{equation}

where

\begin{equation}
\Phi(T_{\rm k}) \equiv  \left ( \frac { 2\,\pi^{1/3} k }  { h c } \right )^{\!3} 
 				   \frac { T_{\rm tr}^2 }  { A_{\rm ul} }
                         		   \frac { F(T_{\rm k}) }   { F(T_{\rm k}) - F(T_{\rm bg}) } 
		     		   \frac { Q(T_{\rm k}) }  { g_{\rm u} }
				   \exp(T_{\rm u}/T_{\rm k}) 
\end{equation}

with cgs-units of K$^{-1}$\,\cmthree\,s. Here, $f_{\rm b}$ is the beam filling factor for the source which may be smaller than the beam ($0 \le f_{\rm b} \le 1$) and $\eta_{\rm mb}$ is the main beam efficiency. $T_{\rm tr} = h \nu/k$ is the transition temperature, $T_{\rm bg}=2.725$\,K is the temperature of the back-ground radiation field, $F(T) \equiv T_{\rm tr}/[{\rm exp}(T_{\rm tr}/T)-1]$ is the quasi-Planck function, $T_{\rm u}$ is the upper level energy in K, $Q(T_{\rm k})$ is the partition function and $g_{\rm u}=(2J+1)$ is the statistical weight of the upper level and the other symbols have their usual meaning.

\subsubsection{\coiso\ and \molh\ column densities}

We limit the discussion to the central core region, where observed \coiso\,(3-2) line intensities of the +3\,\kms\ component are \tadv = 20\,K\,\kms. The upper level energy lies nearly 32\,K above ground. The spontaneous transition probability for the transition is $A_{32} =  2.158\times 10^{-6}\,{\rm s}^{-1}$, the transition temperature is $T_{\rm tr}=15.813$\,K and the statistical weight of the upper level is $g_{\rm u}=7$. Using the collisional rate coefficients of \citet{schinke1985} for collisions with para-\molh, yields critical densities, $n_{\rm crit} \sim A_{32}/\gamma_{32}(T_{\rm k})$, of about $2 \times 10^5$ to $3 \times 10^4$\,\cmthree\ for $T_{\rm k}=10$\,K to 300\,K, respectively (Table\,\ref{col_data}). Therefore, except perhaps for the very lowest temperatures, the condition of LTE should be fulfilled for the \coiso\,(3-2) transition (cf. Sect.\,4.1).

The sizes of the clumps are comparable to the beam size, so that the main beam efficiency, $\eta_{\rm mb}=0.73$, is used for the intensity calibration and we assume here a beam filling factor of unity. For the broad range of temperatures of 10 to 300\,K, the corresponding column densities of \coiso\ are listed in Table\,\ref{col_data}. For an $X$(\coiso)$=1\,{\rm to}\,2 \times 10^{-7}$, the derived \molh\ column densities, on the 20\asec\ scale (2400\,AU), are $N$(\molh)$\,=(1 \pm ^{\,\,3}_{\,\,0.5}) \times 10^{23}$\,\cmtwo. These results are in general agreement with those reported by others \citep{loren1990,motte1998}. Possible opacity corrections to the \coiso\,(3-2) intensity, of the order of $\tau/(1-\exp -\tau)$\,\gapprox\,2, would increase the column density accordingly. The column densities presented in Table\,\ref{col_data} are therefore likely lower limits.

\subsubsection{\moloiso\ column density and \moloiso\ abundance}

The $N_J=2_1-0_1$ transition has the largest Einstein coefficient of the low-lying  \moloiso\ transitions, viz. $A_{21,01} = 1.33 \times 10^{-8}\,{\rm s}^{-1}$ \citep{marechal1997b}. We adopt the coefficients for collisional de-excitation, $\gamma_{21,01}(T_{\rm k})$, which are based on the work by  \citet{bergman1995} and which have been derived  for collisions with He. For  collisions with \molh, these were multiplied by 1.4. Values for temperatures other than 300\,K were obtained by scaling with the square root of the temperature. From Table\,\ref{col_data}, it can be seen that critical densities for the $2_1-0_1$ transition are rather low for a wide range of temperatures ($<1500$\,\cmthree\ above 10\,K). In particular, for the dense core conditions of \roa, where densities are in excess of \powten{5}\,\cmthree\ (Sect.\,4.1), LTE is certainly a valid assumption \citep[see also][]{black1984,marechal1997b}. The temperature of the transition is $T_{\rm u}=T_{\rm tr}=11.228$\,K and the statistical weight of the upper level is $g_{\rm u}=3$. 

In Table\,\ref{col_data}, the results for \coiso\ and \moloiso\ are compared. The ratio $N$(\coiso)/$N$(\moloiso) exceeds unity and increases with decreasing temperature. This ratio could correspond to about half the value of that of the  CO/\molo\ ratio \citep{black1984}. For three cold cores (10 and 15\,K), \citet{fuente1993} determined CO/\molo\,$> 3-7$, limits consistent with, but considerably smaller, than the values displayed in Table\,\ref{col_data}. The effects of a varying C/O ratio in the ISM at column densities (values of the visual extinction \av) as high as those found in \roa\ were explicitly considered in the models by \citet[][see their Fig.\,10]{marechal1997a}. For the \molo\,119\,GHz line, the integrated intensity is $> 10$\,K \kms\ for C/O\,$< 0.4$ when \av\,$> 20$\,mag. In contrast, for similar extinction, the intensity is $< 100$\,mK \kms\ for C/O\,$> 1$. Future observations will likely be able to follow any variation of this ratio in different regions of the ISM \citep[see below and also][]{black1984}.

\begin{table}[ht]
\begin{flushleft}
 \caption{\label{col_data} Column densities of \coiso\ and \moloiso.}
\resizebox{\hsize}{!}{
\begin{tabular}{llllll}
\hline 
\noalign{\smallskip}
$T_{\rm k}$	&   $A_{3,2}/\gamma_{3,2}$ &   $N$(\coiso)		& $A_{21,01}/\gamma_{21,01}$  &  $N$(\moloiso) 	& {\underline{$N$(\coiso)}}  \\   
(K)                   	&  (\cmthree)	   		    &	(\cmtwo)			& (\cmthree)$^a$   			   &  (\cmtwo)$^b$   	& $N$(\moloiso)		   \\ 
\noalign{\smallskip}
\hline
  \hline
  \noalign{\smallskip}
300			&    $3.4 \times 10^{4}$	&$4.9 \times 10^{16}$	& $2.7 \times 10^{2}$	& $<5.6 \times 10^{15}$  & $>9$ \\
100			&    $4.0 \times 10^{4}$	&$2.0 \times 10^{16}$ 	& $4.7 \times 10^{2}$	& $<2.0 \times 10^{15}$  & $>10$ \\
\phantom{1}40	&    $1.7 \times 10^{5}$	&$1.2 \times 10^{16}$	& $7.4 \times 10^{2}$	& $<9.5 \times 10^{14}$  & $>14$ \\
\phantom{1}20	&    $1.7 \times 10^{5}$	&$1.4 \times 10^{16}$	& $1.0 \times 10^{3}$	& $<6.3 \times 10^{14}$  & $>22$ \\
\phantom{1}10	&    $1.7 \times 10^{5}$	&$3.2 \times 10^{16}$	& $1.5 \times 10^{3}$	& $<5.5 \times 10^{14}$  & $>59$ \\
 \noalign{\smallskip}
  \hline
  \end{tabular}
   }
\end{flushleft}
  Notes to the Table: \\
  $^a$ Collision rate coefficient $\gamma(T) = 4.9 \times 10^{-11}\,\,(T/300\,K)^{0.5}\,{\rm cm^3\,s^{-1}}$. \\
  $^b$ Limits on $N$(\moloiso) are $1 \sigma$; $f_{\rm b} = 1.0$ and \tadv\,$< 3.3$\,mK \kms.
  \end{table}

\begin{figure}[ht]
  \resizebox{\hsize}{!}{
  \rotatebox{00}{\includegraphics{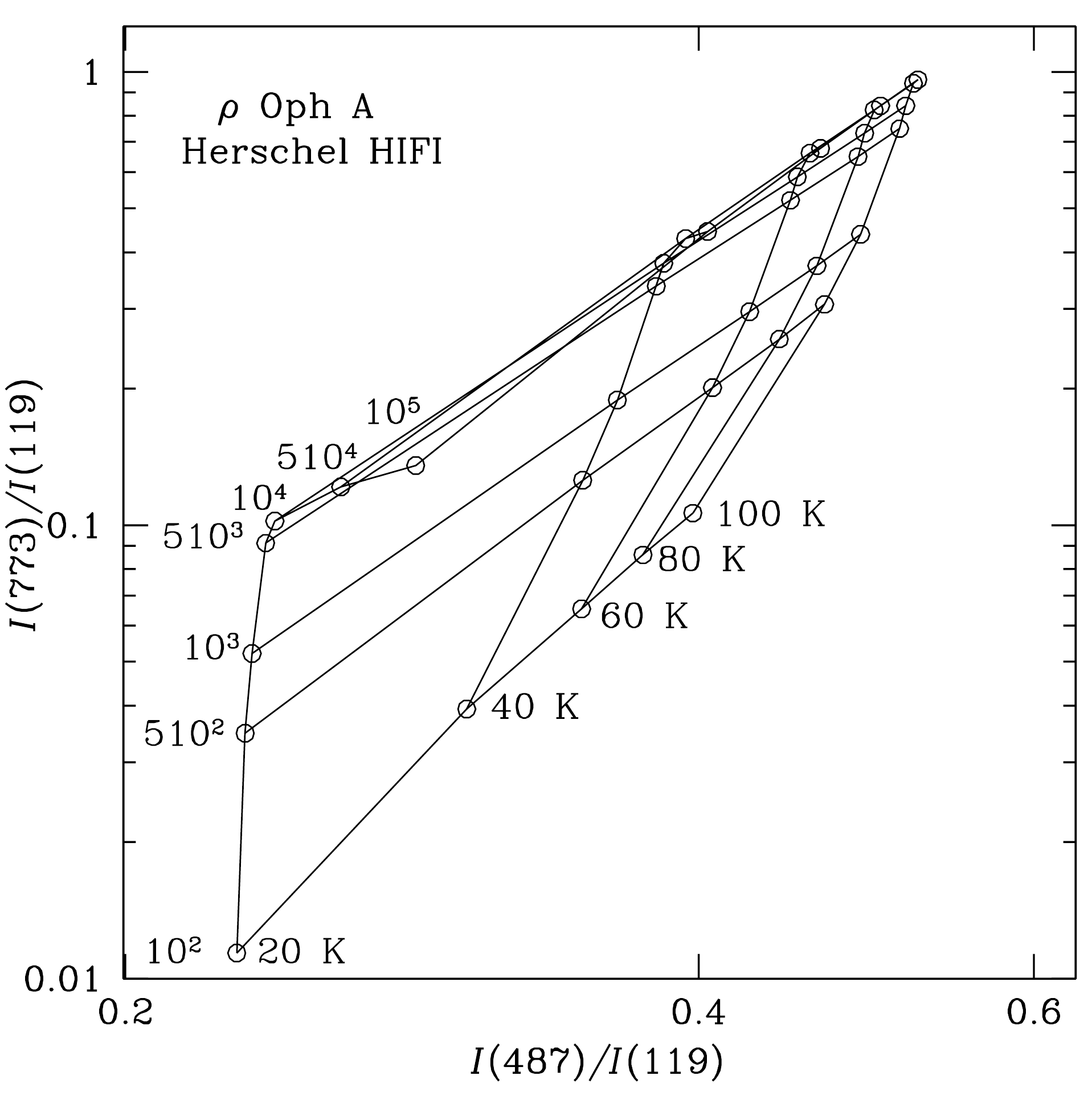}}
  }
  \caption{\molo\ line ratio diagram for the two strongest transitions accessible to HIFI aboard the Herschel Space Observatory. In these multi-transition calculations, radiation from dust was included and LTE was not assumed. Intensity ratios are relative to the $(1_1$ - $1_0)$ 119\,GHz line, which was detected by Odin \citep[$24 \pm 4$\,mK\,\kms\ in a 10\amin\ beam;][]{larsson2007}. These transitions are the $(3_3$ - $1_2)$ 487\,GHz line (44\asec) and the $(5_4$ - $3_4)$ 773\,GHz line (28\asec), respectively. Labels along the graph refer to the density, $n$(\molh) in \cmthree, and the gas temperature, $T_{\rm k}$ in K.}
  \label{o2_line_ratios}
\end{figure}

In the dense cold ($<<100$\,K) regions  of the \roac, the column density of \moloiso\ is lower than $10^{15}$\,\cmtwo\ (Table\,\ref{col_data}) and, hence, the abundance relative to \molh,  $X$(\moloiso)\,$ < 10^{-8}$. Consequently, for the range of 10 to 40\,K and a standard elemental isotopic ratio, the abundance of the primary species of molecular oxygen, $X$(\molo)\,$\sim 500/2 \times X$(\moloiso) \citep{wannier1980}, should be limited to $< 5 \times 10^{-7}  - 2 \times 10^{-6}$, consistent with the Odin result \citep{larsson2007}. We can conclude, therefore, that in the \roac, the molecular oxygen abundance is bounded by $5 \times 10^{-8} \leq X({\rm O_2}) < 2.5 \times 10^{-6}$, where the beam averaged \molo\ column density is \powten{15}\,\cmtwo. If reflecting the fraction of the Odin beam\footnote{Odin observations resulted in a column density of oxygen $N(\rm O_2) = 1\times 10^{15}$\,\cmtwo\ \citep{larsson2007}.  If $\Omega (\rm O_2)=\Omega (\rm C^{18}O)$, the beam filling of the \molo\ source is \powten{-3} to \powten{-2}, i.e. the beam corrected $N(\rm O_2) = 10^{17}-10^{18}$\,\cmtwo. If $N(\rm O^{18}O)$/$N(\rm O_2)=1/250$, then the expected column density of isotopic oxygen is likely within $0.4 \times 10^{15}$\,\cmtwo\, \lapprox\, $N(\rm O^{18}O)$\, \lapprox\, $4 \times 10^{15}$\,\cmtwo\ for a source of size about 20\asec\ to 60\asec.} that is filled by the \molo\ source, its implied size is \lapprox\,$600^{\prime \prime}/\sqrt{250}=38$\asec. This could be well-matched to the 3.5\,m telescope of the Herschel Space Observatory\footnote{\ttfamily http://herschel.esac.esa.int/}, the beam widths of which are 44\asec\ at 487\,GHz, the frequency of the \molo\,($N_J=3_2$ - $1_2$) transition, and 28\asec\ at 773\,GHz for the $(5_4$ - $3_4)$ line (Fig.\,\ref{o2_line_ratios}).

\begin{figure}[ht]
  \resizebox{\hsize}{!}{
   \rotatebox{00}{\includegraphics{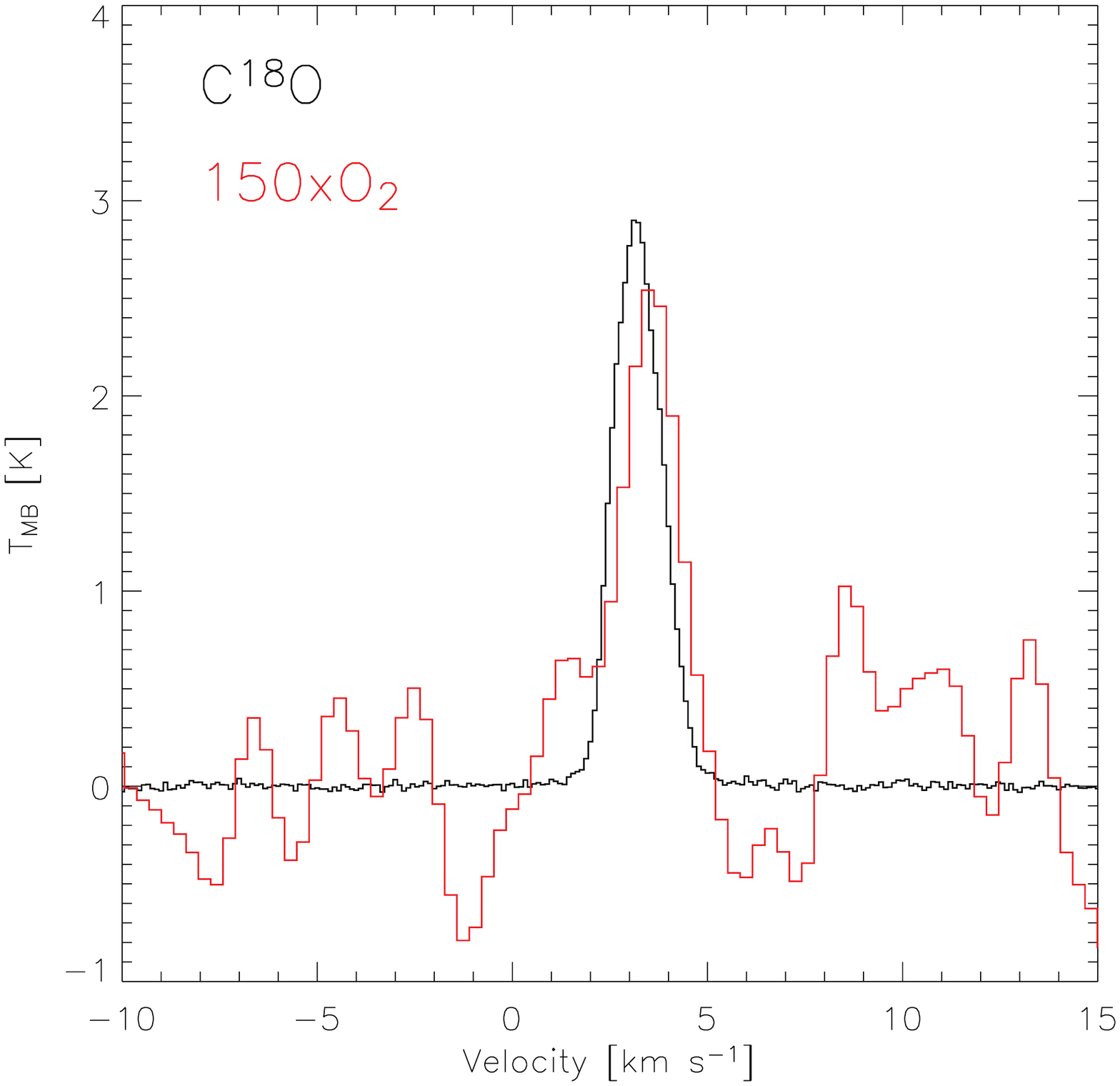}}
  }
  \caption{The line profile of the \coiso\,(3-2) map, after convolution with a 600\asec\ beam, is shown in black and is compared to the scaled \molo\ line of \citet{larsson2007} shown in red. The higher S/N APEX data are spectrally sampled at a higher rate and have also higher resolution. The \coiso\ line intensity is dominated by an extended cloud component at an LSR-velocity seemingly different from that of the P\,3 core and the \molo\ emission (see also Table\,\ref{convol}).}
  \label{convolved}
\end{figure}

\subsection{Nature, location and extent of the \molo\ source}

\subsubsection{Oxygen in the cold ISM}

The capital letter designation of the cores was introduced by \citet{loren1990} for the location of emission peaks in lines of DCO$^+$ in the \roc. Depending on the details of the considered models of the deuteration process, they derived kinetic gas temperatures inside the cores which were always low, in the range 18-23\,K, whereas temperatures in the outer layers were considerably higher. 

In the interstellar medium, it is expected that most of the molecular oxygen is formed by the reaction $$ {\rm O} + {\rm OH}  \to {\rm O}_2 + {\rm H} \;\;\;. $$
The thermal rate coefficient of this reaction has recently been measured in the laboratory down to temperatures of 39 K (Carty et al. 2006), where it remains rapid. In subsequent ab initio theoretical calculations, \citet{xu2007} found a much smaller rate at temperatures below 30\,K and suggested that this might solve the problem of missing O$_2$ in cold interstellar clouds. The low-temperature behaviour of the reaction is also of interest for ultra-cold collisions. \citet{quemener2009} have determined that the reaction still proceeds in the limit of zero temperature with a rate coefficient of approximately $6\times 10^{-12}$ cm$^3$ s$^{-1}$. \citet{quan2008} re-examined the sensitivity of the interstellar O$_2$ abundance to the low-temperature behaviour of the source reaction.

Also a widely favoured explanation for the generally observed paucity of molecular oxygen in the gas phase is depletion of atomic oxygen with subsequent hydrogenation on cold grain surfaces. This scenario seems supported by several observed molecules. For instance, the hydrogenation of CO is predicted to lead to H$_2$CO, CH$_3$OH and subsequently also the deuterated forms of these species \citep{matar2008,fuchs2009}. In order to become observable, these species have to be returned into the gas phase. Indeed, enhanced emission in methanol and doubly deuterated formaldehyde has been observed toward the centre of \roa\ by, respectively, \citet{liseau2003} and Bergman et al. (in preparation). In addition, widespread emission of gas phase \water\ in \roa\ is also observed (Larsson et al., in preparation), a fraction of which may have been similarly produced by the hydrogenation of \molo\ on cold grain surfaces \citep{ioppolo2008}. The equilibrium between adsorption and desorption of molecules would then naturally lead to low levels of both species (as compared and in contrast to pure gas phase chemistry). 

\begin{table}
\begin{flushleft}
 \caption{\label{convol} Gaussian parameters of \coisoiso\,(3-2) and \molo\ (1$_1$-1$_0$) lines.}
\resizebox{\hsize}{!}{
\begin{tabular}{rccll}
\hline 
\noalign{\smallskip}
Molecular				& \vlsr  	& FWHM 	& $T_{\rm peak}$			 & Note \\   
spectral line                     	&  (\kms)   &    (\kms) &  (K)			 &           \\ 
\noalign{\smallskip}
\hline
  \hline
  \noalign{\smallskip}
\molo\,119\,GHz	& $3.5$	& $1.5^{a}$	&$17.4\times 10^{-3}$& observed 10\amin\ beam	\\
\coiso\ 329\,GHz	& $3.2$	& $1.4$	& \phantom{1}2.9	  & convolved 10\amin\ beam	\\
\coiso\ 329\,GHz	& $3.6$	& $1.0$	& 16.5			  & observed 19\asec\ beam, P\,3\\
 \noalign{\smallskip}
  \hline
  \end{tabular}
    }
\end{flushleft}
$^{a}$ The \molo\ line is artifically broadened \citep[see][]{larsson2007}.
\end{table}

\subsubsection{Site and size of the \molo\ source}

Figure\,\ref{convolved} shows the \coiso\,(3-2) spectrum after the convolution of the observed map with a 10\amin\ beam. A Gaussian profile provides a good fit to the observed line, the parameters of which are \tas =2.9\,K, \vlsr =3.20\,\kms\ and \dv =1.45\,\kms\ (Table\,\ref{convol}). This velocity is offset from that of the core P\,3 (3.62\,\kms, cf. also Table\,\ref{13c18o32}) and the intensity is dominated by an extended component. The integrated value is \tadv =4.5\,K\,\kms. Using the temperature assumed by \citet[][i.e. 30\,K]{larsson2007} we obtain the beam averaged column density\footnote{This \coiso\ column density, which represents an average over ten arcminutes, implies an \molh\ column density, $N({\rm H_2}) \sim 2.5 \times 10^{22}$\,\cmtwo, a value which has been derived also by other means \citep[][and references therein]{larsson2007}. At the adopted distance of 120\,pc, this translates into an \molh-mass of the \roa\ cloud of \gapprox\,30\,\msun. Not totally unexpected, most of the mass would be contributed on larger scales \citep[cf., e.g.,][]{motte1998,maruta2009}.}  $N({\rm C^{18}O}) \ge 2.5 \times 10^{15}$\,\cmtwo. The comparison with the Odin result, i.e. $N(\rm O_2) = 1 \times 10^{15}$\,\cmtwo, would indicate that $N$(\coiso)/$N({\rm O_2}) > 1$. A \coiso\ abundance that is larger than that of \molo\ would be difficult to explain and would speak against an extended \molo\ emission region.

The \molo\,119\,GHz Odin line shares the LSR-velocity with that of intensity maxima in \coiso\ and N$_2$H$^+$ \citep[this paper and][]{difrancesco2004,difrancesco2009}. It seems therefore reasonable to identify the location of predominant \molo\ emission with the central parts of the cold core \roa, i.e. the region including P\,2 and P\,3 (SM\,1N and SM\,1, respectively) with a probable extent on the 30\asec\ to 1\amin\ scale. 

\section{Conclusions}

Summarising, we briefly conclude the following:

\begin{itemize}
\item[$\bullet$] \coiso\,(3-2) mapping observations with APEX of a $10^{\prime} \times 5^{\prime}$ region in \roa\ have revealed a complex radial velocity field. The central $200^{\prime \prime} \times 200^{\prime \prime}$ have been spatially sampled at the Nyquist frequency.
\item[$\bullet$] The \vlsr\ of the \molo\,119\,GHz line appears confined to a particular region (SM\,1), which is also a prime emitter in \coiso\ and N$_2$H$^+$.
\item[$\bullet$] The observation of \moloiso\ toward SM\,1 (P\,3) and SM\,1N (P\,2) resulted in upper limits. Combined with the \coiso\ data, this leads to a ratio of $N$(\coiso) to $N$(\moloiso) much larger than unity.
\item[$\bullet$] From the \molo\ and \moloiso\ observations we infer an \molo\ abundance $5 \times 10^{-7} < {\rm O_2} \le 2.5 \times 10^{-6}$.
\item[$\bullet$] The \molo\ source is likely relatively compact, within a factor of a few of 1 arcminute, and should become readily detectable by upcoming Herschel HIFI observations.
\end{itemize}

\acknowledgement{We wish to thank Cathy Horellou and Daniel Johansson for making part of the APEX observations.}

\end{document}